# Broadband Emission via a Photon Avalanche in a Lanthanide-Trimesic Acid Metal-Organic Framework

*Hadar Nasi,[1][†] Miri Kazes,[1][†]* Michal Leskes,[1] Hagai Cohen,[2] Ayelet Vilan,[2] Linda J. W. Shimon,[2] Ifat Kaplan-Ashiri,[2] Michal Lahav,[1] Dan Oron,[1]* Maria Chiara di Gregorio[3]*

1 Department of Molecular Chemistry and Materials Science, Weizmann Institute of Science, 7610001 Rehovot, Israel.

2 Department of Chemical Research Support, Weizmann Institute of Science, 7610001 Rehovot, Israel.

3 Department of Chemistry, Sapienza University of Rome, P.le A. Moro 5, 00185 Rome, Italy.

[†]These authors contributed equally to this work

*Corresponding authors

Email: mariachiara.digregorio@uniroma1.it

dan.oron@weizmann.ac.il

miri.kazes@weizmann.ac.il

**Abstract**

Infrared-triggered photon upconversion in porous materials presents intriguing prospects for combined functionalities such as molecular sponge, energy harvesting and conversion functionalities. Metal-organic frameworks (MOFs) are one of the most versatile classes of porous crystals. So far only two-photon upconverting processes have been realized in MOFs both by ligand based triplet-triplet annihilation and directly in lanthanide ions. Here we report on $Yb^{3+}/Er^{3+}$-trimesate-based MOFs that exhibit photon avalanche (PA) characteristics. The PA process conventionally occurs through cross-relaxation within the lanthanide emitter manifold. In contrast, here PA proceeds in the organic molecule part and relies on a cooperative process, involving multiple emission centers. The IR photons are first absorbed and upconverted into high energy electronic population by the action of the lanthanide ions ($Yb^{3+}$ and $Er^{3+}$ are the sensitizer and the activator, respectively). Subsequently, the electrons are funneled into electronically coupled triplet states of the trimesate ligand, enabling accumulation in the organic matrix. This reservoir acts as source for a highly nonlinear spectrally broadband emission, arising mainly from ligand triplet states. The nonlinearity factor is comparable with the well-established PA inorganic nanoparticles. We prove that the PA is strongly related to the degree of crystallinity of the MOF: not well-formed frameworks support only the characteristic $Er^{3+}$ emission with only linear increase as a function of the excitation power. Our work paves a path towards vastly expanding the range of materials exhibiting PA, well beyond a limited set of lanthanide ions. Moreover, it provides a path for much broader control of the PA emission characteristics.

**Introduction**

Upconversion luminescence (UCL) is a nonlinear process where two or more lower energy photons are absorbed and converted into a high-energy emitted photon. Accomplishing such an Anti-stokes shift in the NIR - visible window has broad applicative interest in light harvesting, photocatalysis, bio-imaging, theranostics and sensors.[1–5] As UCL requires long-lived intermediate states, it is usually achieved by one of two distinct strategies: The first is based on parity-protected states in trivalent rare earth ions as intermediate states, and the emission is due to allowed transitions from higher states. Upconversion through trivalent rare earth ions typically results in sharp emission lines which can be associated with a particular atomic transition. The second is based on spin-protected molecular triplets as intermediate states and on a triplet-triplet annihilation process (TTA) for excitation of the emitting singlet state. Emission from TTA thus exhibits characteristics associated with molecular fluorescence such as relatively broadband emission.[6–9]

Rare earth-based UCL typically arise from either a single-ion or an ion-pair system. In the former case the mechanism of upconversion is based on excited state absorption (ESA) where high excited states are populated by consecutive absorption of photons resonating with the involved electronic transitions. In the latter case this is complemented by the interplay of an absorbing (sensitizer) and emitter species that co-work through mechanisms of energy transfer (ETU).[10–12] In recent years, properly designed lanthanide based materials have been shown to exhibit the highly nonlinear upconverted emission termed photon avalanche (PA). The PA process typically involves a cycle of efficient excited state absorption (ESA) followed by energy cross relaxation (CR) between neighboring ions. This cooperative action initiates a looping mechanism which enhances the population of the intermediate level from which ESA takes place. The experimental hallmark of PA is the presence of an excitation power threshold upon which the intensity of the UCL increases in a highly nonlinear fashion. [1–3,5,13,14]

Although rare-earth based UCL phenomena were first observed in bulk solids and solid fibers, a significant branch of recent research has focused on the development of UCL nanoparticles where the rare earth ions are introduced as dopants within a dielectric lattice. The nanoscale size reduction opened the field to nano-applications. However, these synthetic processes are often quite laborious, requiring formation of multi-shell architectures.[13,15–17]

In 2011 Piguet et al. demonstrated that rare-earth based UCL can also occur in discrete molecular entities such as metal-organic complexes.[18,19] The incorporation of both sensitizers and activators within the same molecule structure facilitated energy transfer processes, offering better control over material design and thus UCL pathways compared to UCL nanoparticles.[20–23] Nevertheless, intrinsic problems as multiphonon relaxation through the high-energy vibrational states of the ligand and short lifetime of the excited relays render this mechanism inefficient. To the best of our knowledge, none of these past studies have observed PA in metal-organic complexes.

In parallel with these well-explored classes of UCL materials, increasing attention has recently been directed towards MOFs as UCL materials. MOFs are porous crystals formed by the assembly of metal ions and organic molecules through coordination bonds. They have been extensively studied over the past 35 years for their excellent adsorption and catalytic properties. Combining the intrinsic material porosity with optical properties presents an appealing perspective for many applicative purposes, such as photocatalysis, sensing and phototherapy. However, design and study of the optical properties of MOFs have emerged as prominent research directions only in the last decade.[24–27] Recent studies have shown that the organic linker's triplet state can support either downconversion or upconversion luminescence. For example, the organic linker can harvest incident light and transfer it to the metal ions via triplet states by the antenna effect.[28,29] MOF upconversion has been realized by designing suitable ligands for triplet-triplet annihilation.[30–36] Other upconverting-active MOFs rely instead on formation of heterostructures with inorganics nanoparticles as the UCL materials, while the MOF serves another function such as ion release or biocompatibility. [37-39]

lanthanide-based UCL MOFs such as $Er^{3+}/Y^{3+}$-Benzene-1,3,5-tricarboxylic acid (Trimesic acid, BTC) and $Y^{3+}/Er^{3+}/Yb^{3+}$-Benzene-1,4-dicarboxylic acid (BDC) are limited in number, though the synthetic aspects appear to be more straightforward than their fully inorganic counterparts.[40,41–48] Surprisingly, reports on MOFs with identical crystal packing and building blocks exhibit variations in their UCL spectra.[42,44] We believe that such variance in the state of art is hint of an overlooked versatility of the MOF matrix to variegate UCL upon slight variation of the reaction parameters. A deeper understanding of the synthetic-structural-optical relationship is needed for a more conscious exploitation of the UCL potential of MOFs.

Here we investigate MOFs formed by trimesic acid (BTC) and the $Er^{3+}/Yb^{3+}$ ion-pair which is the most thermally stable and investigated MOF for upconversion.[44,48] However, we show that under certain conditions this system exhibits a previously unobserved broadband PA emission. As such, the PA emission clearly involves the organic component and exhibits complex dynamics on multiple time scales ranging from milliseconds to seconds. We show that this broad band PA emission involves energy transfer from the lanthanide ions to the BTC organic linker and that the emission is commensurate with characteristic emission from the organic linker triplet manifold. Moreover, PA seems to involve cooperative action of lanthanide ions and several adjacent organic linkers. Interestingly, one key parameter controlling the presence and efficiency of the observed PA turns out to be the degree of crystallinity of the MOF.

**Results**

**PA emission characterization and dynamics**

Micro and nano sized $Yb^{3+}/Er^{3+}$ BTC MOFs samples were synthesized under different reaction times: 1 day and 3 days (see **Methods** section and **SI** for full details). In the following text the differently synthesized MOFs will be named by acronyms formed by an initial part specifying the size range of the structures, **nMOF** and **µMOF,** and a final part referring to the reaction time (**-1d** and **-3d** for 1 day and 3 days MOF synthesis, respectively). **µMOF-1d, µMOF-3d** and **nMOF-3d** show a spectrally broad highly nonlinear UCL emission at increased excitation power (which we associate with PA) whereas **nMOF-1d** exhibits only the characteristic $Er^{3+}$ UCL. As a representative PA structure, **µMOF-3d** (SEM image shown in **Figure 1a**) emission properties are discussed below. Analogous data and analysis are reported in the SI for PA in **µMOF-1d** and **nMOF-3d**.

UCL spectra at different excitation powers of **µMOF-3d** are presented in **Figure 1b** and **Figure 1c**. The spectra at low excitation powers show peaks ascribable to the typical $Er^{3+}$ optical transitions: 522 nm (blue line), 539 nm (cyan line), 554 nm (green line), 564 nm (dark green line) emission lines that can be assigned to the $^4H_{11/2}, ^4S_{3/2} \rightarrow ^4I_{15/2}$ transitions, and 654 nm emission (red line) assigned to $^4F_{9/2} \rightarrow ^4I_{15/2}$.[49] Interestingly, after an initial rise in UCL, the emission is quenched with increased excitation power. Remarkably, beyond an excitation power threshold of 25 kW/cm², a broad emission emerges. To demonstrate these changes more clearly, representative spectra at different excitation powers are presented in **Figure 1c**.

The UCL intensity as a function of excitation power at various wavelength across the spectrum, presented is **Figure 1c**, shows three distinguishable regimes: i) Typical $Er^{3+}$ emission lines that increase in intensity with excitation power up to ~10 kW/cm²; ii) a gradual quenching of $Er^{3+}$ emission between 10-25 kW/cm²; and iii) a highly nonlinear increase in intensity of three orders of magnitude from an onset of 25 kW/cm² accompanied by the emergence of a broad emission centered at ~550nm with a full width at half-maximum (FWHM) of ~100nm. The steep increase in emission intensity beyond a certain excitation power threshold is an indication for the emergence of PA. Following the PA model,[13,14,50] the emission intensity at threshold was fitted by the following power law relation $I \propto P^n$ where $I$ is the emission peak intensity, $P$ is the power density and $n$ is typically termed the nonlinearity factor. A value of $n$ = 14 was inferred (black line in **Figure 1c**). This nonlinearity is comparable with recently reported PA UCL from inorganic nanoparticles. Interestingly, the threshold power density for PA, 25kW/cm²,is also comparable.[2,13,14,51] The intensity increase associated with the PA is also observed at 580 and 600 nm (yellow and orange markers, respectively), where no $Er^{3+}$ atomic lines exist. Moreover, to the best of our knowledge, the pre-avalanche quenching behavior described in point ii), indicating depletion of the excitation in the lanthanide system, has no analogues in previously reported PA mechanisms, which exclusively rely on dynamics within the manifolds of lanthanide ions.

The PA process is reversible with respect to excitation power, as demonstrated by the measurements performed during sequential increase and decrease in the excitation intensity (**Figure S2**). This behavior is consistently observed across multiple sample batches and different locations within the same sample (**Figure S3**).

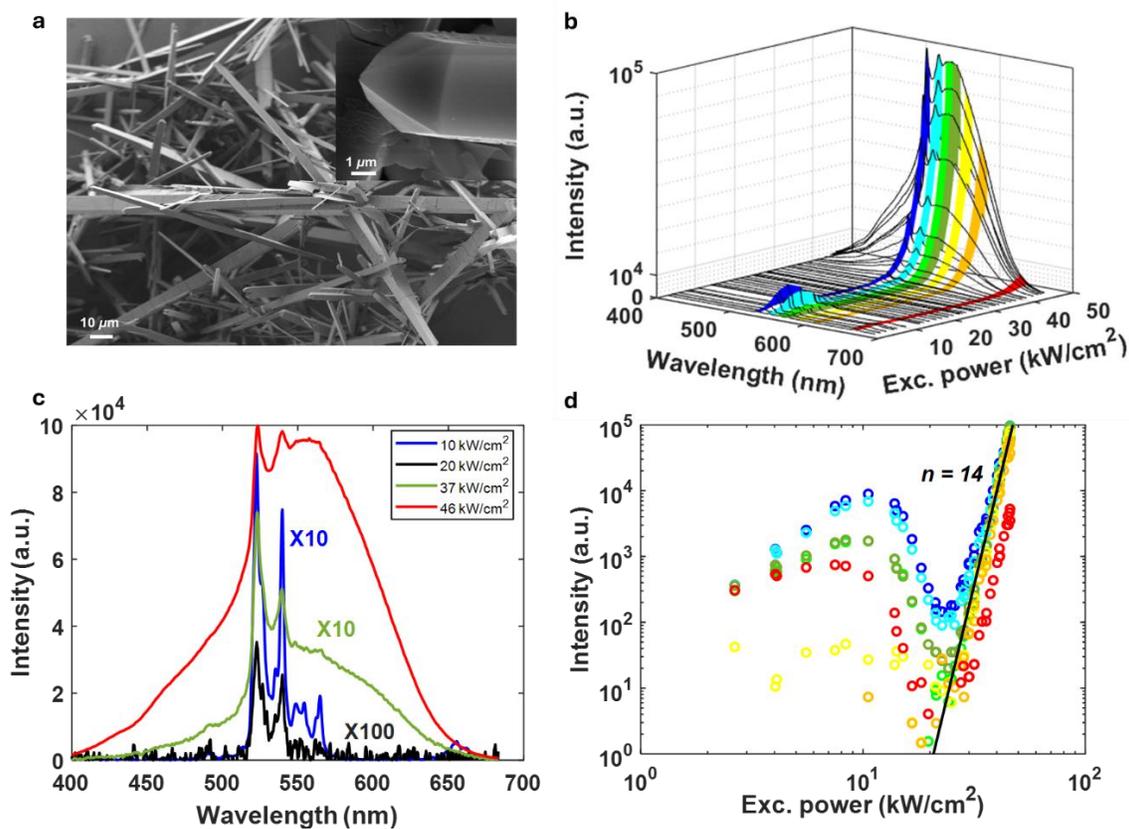

**Figure 1:** Upconversion luminescence (UCL) measurements of **μMOF-3d** as a function of excitation power. a) Scanning Electron Microscopy (SEM) image of **μMOF-3d**. b) UCL spectra for increased excitation powers. c) Representative UCL spectra highlighting the $Er^{3+}$ emission peaks, along with the broad emission that emerges at higher excitation powers. d) UCL intensity as a function of excitation power for different $Er^{3+}$ emission peak positions: 522 nm (blue), 539 nm (cyan), 554 nm (light green), 564 nm (dark green), and 654 nm (red). Notably, a steep increase in intensity is also observed at 580 nm (yellow) and 600 nm (orange), wavelengths at which no $Er^{3+}$ optical transitions exist. A power-law fit of the data at 564 nm, with a nonlinear factor of n=14, is plotted in black.

The PA process shows maximum efficiency when the excitation wavelength is in resonance with the $Yb^{3+}$ 980 nm transition line (**Figure S4**). This fact further distinguishes our material from photon-avalanching inorganic nanoparticles. Indeed, PA of lanthanides in an inorganic matrix usually relies on off-resonant excitation of the sensitizers to increase the ratio of excited state to ground state absorption rates, thereby promoting cross-relaxation.[3,5]

Transient emission measurements were taken by modulating the excitation beam using an optical chopper at a repetition rate of circa 2 Hz and are presented in **Figure 2a** along with the integrated emission spectrum under similar excitation conditions, presented in **Figure 2b**. As can be seen from **Figure 2a**, the emission intensity exhibits nontrivial dynamics ranging from millisecond to a second time scale. At an excitation power below the PA threshold (20 kW/cm$^2$) where the characteristic Er$^{3+}$ emission spectrum is obtained, there is a fast rise in emission followed by a two-component decay to a steady-state level. The fast decay component is of the order of 1 ms, and the slow component of ~60 msec (**Figures 2a,2b**, blue lines, respectively). At the PA threshold (28 kW/cm$^2$, black lines) the spectrum still resembles that of the Er$^{3+}$ system, but with significantly reduced intensity. However, the second decay component shortens 40 ms, bringing the intensity to nearly complete depletion within 200 msec. At 41 kW/cm$^2$, already above the PA threshold (green lines), broadening of the spectrum begins to show. This is accompanied by a further shortening of the second decay component to 15 ms, and it is followed by a slow (circa 100ms) rise to the steady-state PL. At the maximum available power of 46 kW/cm$^2$ (red lines) broad PA is fully observed, the (near) steady state PL is stronger than that observed from the initial Er$^{3+}$ emission maximum.

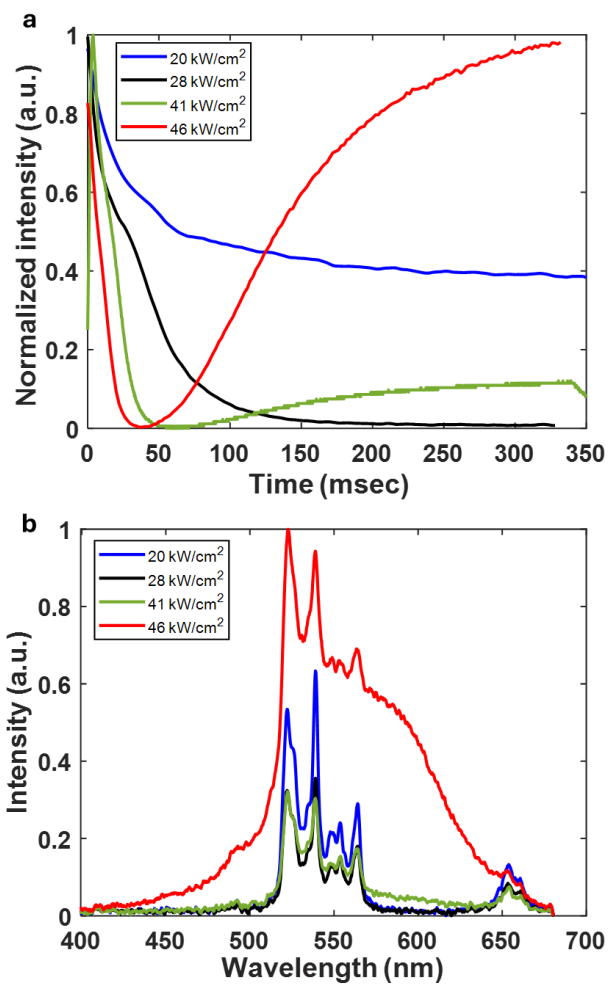

**Figure 2:** Transient UPL measurements at 2 Hz modulation of the pump beam. a) UCL transients for different excitation powers. b) UCL spectra taken for different excitation powers.

**Analysis of the origin of the broadband emission**

To understand the origin of the broadband emission, reference samples of $Er^{3+}$-only and $Yb^{3+}$-only based trimesate MOFs were synthesized and measured. In both cases, the emission was unstable and sporadic across the sample area, and emerging only under the highest available excitation power. For the pure $Er^{3+}$ sample, two different spectra appeared arbitrarily. One corresponds to the characteristic $Er^{3+}$ emission, as shown in **Figure 3a** (green curve). While the other is a broad emission spectrum, presented in Figure 3b (green curve). The $Yb^{3+}$-only sample exhibited merely the broad emission as shown in **Figure 3b** (blue curve). The broad emission spectra of the two reference samples closely resemble the broadband profile observed in the Yb/Er MOF (**Figure 3b**, red curve). The similarity between all three samples strongly suggests that the organic component is responsible for the observed broadband emission. However, no emission was detected from pure trimesic acid powder, highlighting the importance of the charged state and possibly the three-dimensional organization in enabling the emission process. To complement and support these findings, we performed cathodoluminescence (CL) measurements on both the MOF samples and pure trimesic acid. The CL spectrum of pure trimesic acid, shown in **Figure 3b** (black curve), exhibits a further broadened emission having a pronounced peak at 560 nm and a distinguishable shoulder between 400 and 500 nm. The BTC CL spectrum can be deconvolved to three Gaussians centered at 445 nm, 564 nm and 571 nm (**Figure S5**). The CL emission spectra of the MOF samples (**Figure S6**) resemble that of the BTC spectrum but also contain signs of atomic $Er^{3+}$ emission.

Light emission from trimesate, known to be phosphorescent at room temperature, has been previously reported in the literature. Phosphorescence spectra present three bands at about 410 nm, 560 nm and 640 nm with relative intensities, strongly dependent on the excitation wavelength.[52] Similar emission spectra for trimesate triplet state have been derived also from Gd-trimesate complexes.[53] Spectral deconvolution of the PA **μMOF-3d** and pure $Er^{3+}$- and $Yb^{3+}$- trimesate MOFs enables to retrieve three significant components centered at 541 nm, 558 nm and 603 nm (**Figure 3c**), with the 560 nm being the dominant contribution. In the **μMOF-3d** emission fit, a contribution from the $Er^{3+}$ optical transitions centered at 523 nm and 536 nm could also be accounted while the rest of the $Er^{3+}$ lines are suppressed. These $Er^{3+}$ components are small in comparison to the contribution form the emission assigned to the BTC, suggesting that PA is proceeding within the BTC manifold.

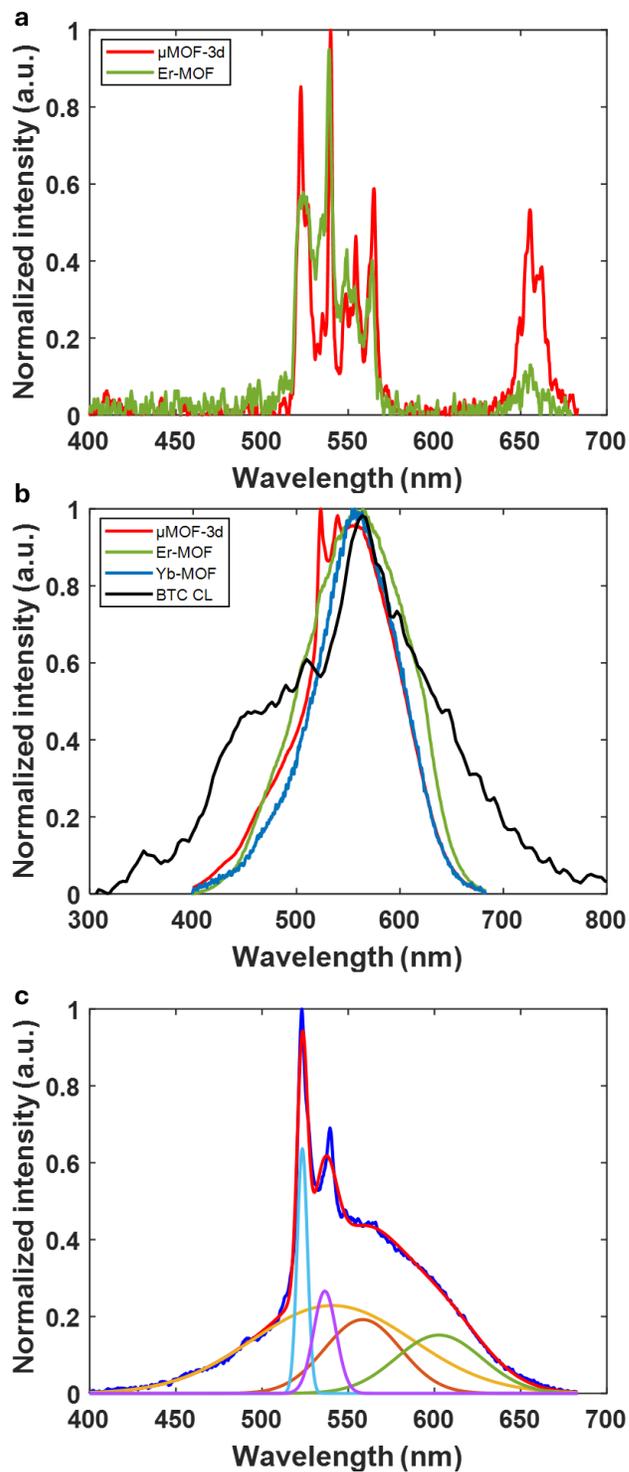

**Figure 3:** Upconversion luminescence (UCL) measurements of $Yb^{3+}$-only and $Er^{3+}$-only MOFs. a) UCL spectra of $Er^{3+}$-only (green line) and $Yb^{3+}/Er^{3+}$ (red line) MOFs. Both spectra show the typical $Er^{3+}$ optical transitions. b) UCL spectra of $Er^{3+}$-only (green line), $Yb^{3+}$-only (blue line) and $Yb^{3+}/Er^{3+}$ (red line) MOFs. A cathodoluminescence (CL) spectrum of pure

trimesic acid is presented in black. All spectra display a broad emission centered around 560 nm. c) A multi-Gaussian fit of the Yb$^{3+}$/Er$^{3+}$ MOF.

**Effect of cooperativity**

Further measurements were carried out on **nMOF-1d** showing no PA. Power-dependent UCL spectra are presented in **Figure 4a** showing only the characteristic Er$^{3+}$ optical transitions, with the broadband emission entirely absent. The power law exponent dependence of Er$^{3+}$ emission on excitation power is a modest ~3 in this sample, highlighting the dramatic difference with the above-described PA mechanism. (**Figure 4b**).

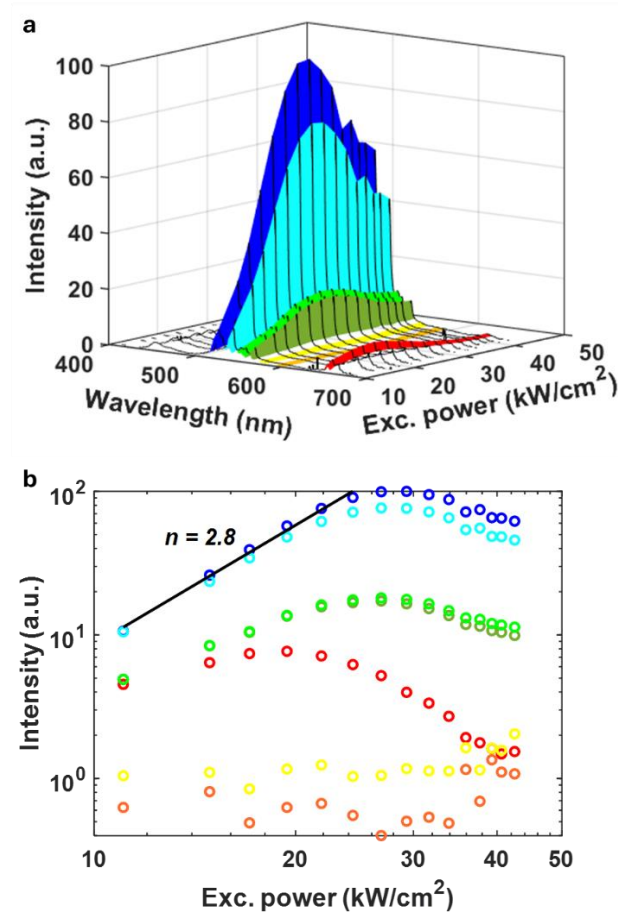

**Figure 4:** Upconversion luminescence (UCL) of **nMOF-1d**. a) UCL spectra of **nMOF-1d** at different excitation powers. The typical Er$^{3+}$ emission peaks are seen at 522 nm (blue), 539 nm (cyan), 554 nm (light green), 564 nm (dark green) and 654 nm (red). b) Intensity as a function of excitation power at the different wavelengths marked as above. The power dependence of the 522 nm emission follows a power law with an exponent of n = 2.8.

The absence of PA in the **nMOF-1d** sample likely arises from structural or compositional differences. To investigate this, we utilized multiple spectroscopic techniques, including single crystal X-ray diffraction (SCXD, **Figure S7**), powder X-ray diffraction (PXRD, **Figure S8**), thermal gravimetric analysis (TGA, **Figure S9**) and FTIR (**Figure S10**). However, none of these methods revealed significant differences between samples exhibiting PA and non-PA **nMOF-1d** sample.

The only clear indication of structural variance arises from $^1$H magic angle spinning (MAS) solid state NMR spectroscopy which is suitable for detecting local variations around the metal ion environment. $^1$H MAS NMR isotropic spectra of the **nMOF-1d** and **nMOF-3d**, shown in **Figure 5** (see **Figure S11** for full spectra and details), both display three main resonances at about 10 ppm, 1 ppm and -11 ppm. These can be assigned to BTC ligands which are the main source of hydrogen in the sample. As the ligands are bound to the paramagnetic lanthanide ions, in addition to the chemical shift, their resonance frequency is shifted through bonds and space interaction with the f electrons. Additional resonances are observed in **nMOF-1d** sample at about -66 ppm and +50 ppm. These resonances, absent in the **nMOF-3d** sample, may arise from residual solvent molecules around the lanthanide ion. The large shift indicates these solvent molecules are in very close proximity to the lanthanide. To examine whether these resonances arise from solvent environments, the samples were soaked in $D_2O$ followed by vacuum drying. After treatment, the two resonances observed in the **nMOF-1d** at high shifts disappeared, supporting their assignment to solvent molecules weakly coordinated to the lanthanide ions. The main resonances around +10 ppm, +1 ppm and -11 ppm remained following this treatment, in accordance with their assignment as ligand environments. Some variation in the ligand resonances position, width and relative intensities can be observed after $D_2O$ soaking, likely due to changes in the magnetic resonance properties of the lanthanide ions. We therefore postulate that the presence of these additional weakly bound solvent molecules inhibits the PA process, possibly by providing rapid nonradiative vibrational relaxation pathways.

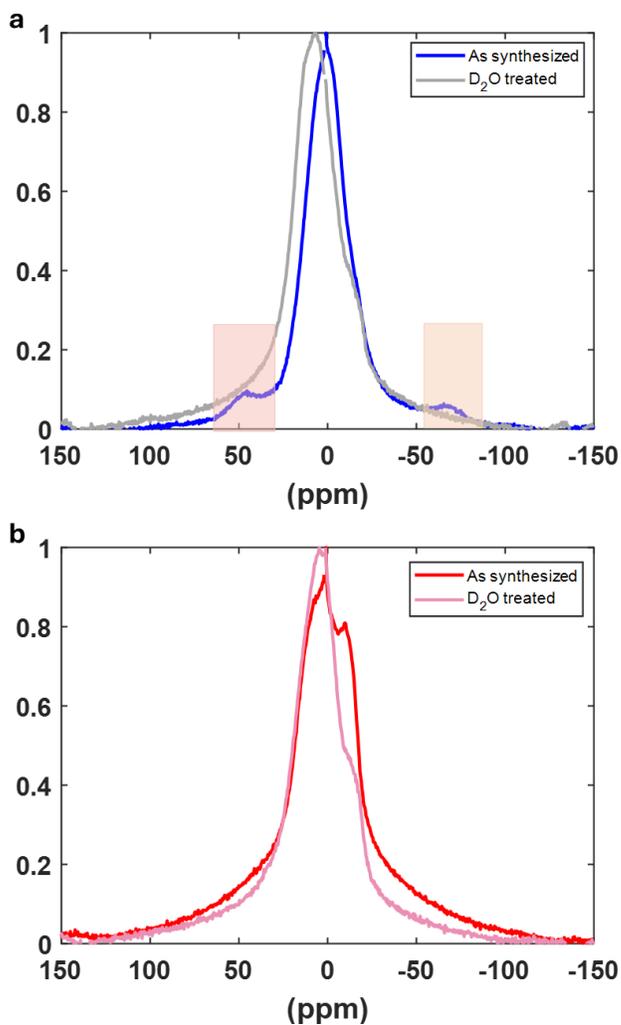

**Figure 5.** Isotropic-$^1$H magic angle spinning (MAS) solid-state NMR spectra of a) **nMOF-1d,** and b) **nMOF-3d** both treated with solvent exchange/drying processes and subsequent exchange with deuterated water.

**Discussion**

We have shown that Yb/Er-based MOFs can support highly nonlinear PA, comparable to that observed in inorganic nanoparticles. [2,13,14,51] The two general requirements for PA is a large ratio of the excited state absorption to the ground state absorption and a cross-relaxation mechanism which exponentially increases the population of the intermediate state.[5] The former is practically automatically fulfilled in our system since the ground state transition of the organic is phosphorescent. We cannot clearly identify the higher states involved in the latter (or even the exact mechanism of its excitation, be it by one-photon or by two-photon absorption), but the fact that PA is observed in either $Er^{3+}$-only or $Yb^{3+}$-only MOFs strongly connects it to the organic triplet manifold. The Er/Yb system appears to significantly facilitate excitation of the ligand triplet state through resonant energy transfer from the $Er^{3+}$ manifold.

Typically, PA is characterized by a power law dependence having a linear pre-avalanche region, followed by the PA steeply increase at a threshold power and subsequent saturation region. [1–5] In our system, PA is truly a threshold phenomenon, which involves a different transition than the ones which initialize the excited state population in the organic ligands. The quenched $Er^{3+}$ emission in the pre-avalanche region is indicative of excitation transfer from a high state in the $Er^{3+}$ manifold to the BTC, either to the first excited singlet state (as depicted in the scheme, **Figure 6**) or directly to the triplet states and likely followed by a state filling of the triplet states. Subsequently, the PA emission emerges from the ligand triplet states, consistent with its broadband feature.

The fact that other structurally analogous reported MOFs have not shown PA,[44,48] as well as **nMOF-1d**, indicates that this PA process likely goes beyond the single coordination node and occurs when synthetic procedures enable a longer range electronic communication. This suggests that not only the intrinsic hybrid nature of the MOF but also its reticular collective property support this phenomenon. Likely, due to this reticular behavior we cannot exclude non radiative pathways competing with emission and depleting the triplet state population. A hint supporting this hypothesis is the observation of highly unstable emission in the $Er^{3+}$-only MOF which is only detectable at high excitation powers.

Indirect support for the above arguments is gained from our XPS analysis, see the SI file: Shake-up satellites next to the O 1s, C 1s, Yb 4d and Er 4d core lines provide several independent indications for cross-talk between the Yb and Er atomic sites, as well as between the aromatic rings and their COO-M side groups. First, the effect of aromatic-ring excitations on the COO side groups, possibly also on the metallic sites, is clearly seen in the 6-7 eV regime. Second, sensitivity of the Yb and Er shake-up features to the identity of neighboring metallic sites can be seen. These spectral observations emerge from high-energy excitations at very short time scales ($10^{-17}$ s). However, in spite of clear differences in this respect, the up-conversion processes discussed here shares a feature common to the shake-up processes: both reflect a non-local response of the electronic configuration to atomic-scale electron excitations. Therefore, the XPS observations support our proposed interpretation of the optical up-conversion mechanism, where involvement of the organic skeleton in atomic-originated electron transitions can become inseparable and even dominant.

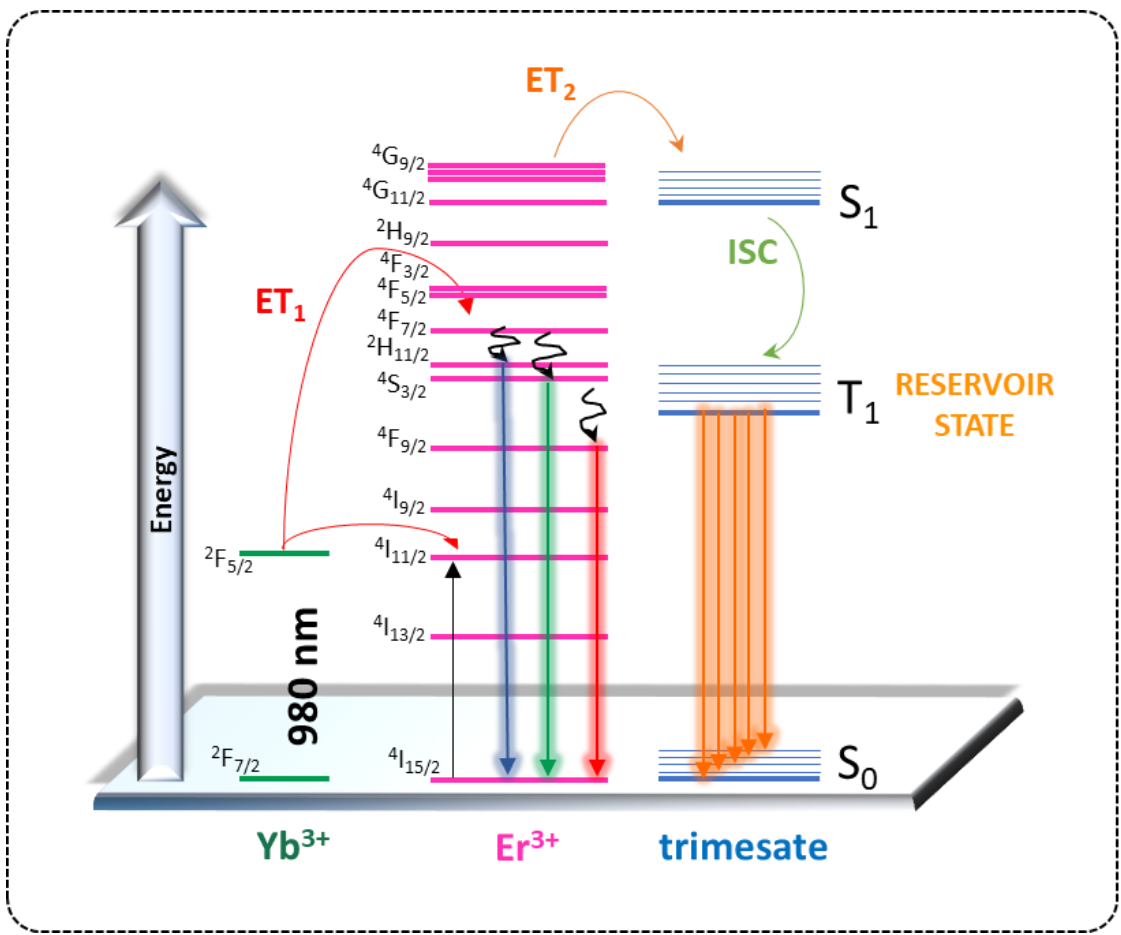

**Figure 6:** Schematic illustration of the photon avalanche (PA) mechanism in $Yb^{3+}/Er^{3+}$-BTC MOF. Upconversion (UC) through the $Yb^{3+}$ excited state is followed by energy transfer (ET) to neighboring $Er^{3+}$ centers, resulting in emission through the $Er^{3+}$ electronic manifold. Concurrently, ET occurs from the blue and green energy levels of $Er^{3+}$ to the BTC either the singlet or directly to the triplet states. Where the BTC serves has an energy reservoir by merit of its long-lived triplet state. This process is evidenced by the gradual quenching of $Er^{3+}$ emission peaks. Moreover, the MOF reticular nature allows for energy migration through the BTC network across extended lattice sites. With increased excitation power, more electrons populate the BTC triplet state giving rise to PA emission from the BTC triplet state manifold.

In conclusion we demonstrate PA emission from $Yb^{3+}/Er^{3+}$-BTC MOFs, resulting in broadband emission assigned to the organic ligand triplet states. This is mediated by excitation transfer from the $Er^{3+}$ electronic states to the ligand triplet state, which serves as an energy reservoir. The phenomenon seems to be collective where, if the crystal framework is defective, the PA is not activated. Our results offer a new perspective to enlarge the PA particle design and emission properties, extending the set of designing parameters beyond the common lanthanide gamut. High order architecture with laborious multi step synthesis is required in PA inorganic nanoparticles. [13,15–17] Here, we show that MOFs can enable analogue optical properties upon simple synthetic approches. Furthermore, the adaptive coordination nodes along with the crystal structure design can potentially serve the electronic and optical properties manipulation.

**Methods**

Synthesis of Yb/Er-BTC MOFs

Yb/Er-BTC MOFs were synthesized by dissolving BTC (0.1 mmol), Yb(NO$_3$)$_3$•5H$_2$O (0.094 mmol) and ErCl$_3$•6H$_2$O (0.006 mmol) in water/DMF (4 ml/8 ml) and subsequently reacting the solution at 60°C for 1 day or 3 days. The chosen Yb:Er mole ratio (94:6) is commonly used in upconversion systems.[54] The synthesis resulted into micro crystals (from about 60 to 100 μm in length and about 3 μm in cross-section, see size distribution in SI Table S1, having rectangular cross-section and pyramidal termini. In addition, nanocrystals were also produced by an analogue synthesis where sodium acetate (0.1 mmol) was added to the building block solution (length about 0.6 μm and cross-section of about 0.3 μm, see size distribution in SI, Table S1). The addition of sodium acetate induces already at room temperature the formation of a fine dispersion. Similarly to what described above, such sample is subsequently heated up at 60 °C for 1 day or 3 days, forming nanostructures, morphologically similar to those formed without sodium acetate, as imaged by SEM microscopy presented in Figure 1a and Figure S1.

Upconversion luminescence (UCL) measurements

UCL measurements were conducted using pulsed laser excitation from a Ti:Sapphire ultrafast laser oscillator delivering 100 fs pulses centered at 980 nm at a repetition rate of 80 MHz (spectra physics Mai Tai). The excitation wavelength is in resonant with the Yb$^{3+}$ $^2F_{7/2} \rightarrow {}^2F_{5/2}$ transition. MOFs in powder form were pressed onto a glass coverslip and mounted at a ~45°relative to the detection line. Transient emission was measured by modulating the excitation beam using a chopper on the excitation line and collecting the emission into a PMT detector, coupled to an oscilloscope. All experiments were carried out at room temperature and in ambient conditions. UCL measurements were conducted as a function of the laser power excitation.

**Acknowledgment**

We thank Prof. Omer Yaffe, Dr. Iddo Pinkas and Yisrael Shlomo Rand for their efforts in the material carachterization.

The crystallographic data has been submitted in the database Cambridge Crystallographic Data Centre. The CCDC numbers are 2452337-2452340.